\documentclass[sigconf]{acmart}
\AtBeginDocument{%
  }

\usepackage{tabularx}
\usepackage{multirow}
\usepackage{balance}
\usepackage{natbib}
\copyrightyear{2025}
\acmYear{2025}
\setcopyright{cc}
\setcctype{by}
\acmConference[ICMR '25]{Proceedings of the 2025 International Conference on
Multimedia Retrieval}{June 30-July 3, 2025}{Chicago, IL, USA}
\acmBooktitle{Proceedings of the 2025 International Conference on Multimedia
Retrieval (ICMR '25), June 30-July 3, 2025, Chicago, IL, USA}
\acmDOI{10.1145/3731715.3733482}
\acmISBN{979-8-4007-1877-9/2025/06}

\begin{document}

\title{Latent Sensor Fusion: Multimedia Learning of Physiological Signals for Resource-Constrained Devices}

\author{Abdullah Ahmed}
\email{amahmed@umass.edu}
\orcid{0000-0002-9048-4242}
\affiliation{%
  \institution{University of Massachusetts Amherst}
  \city{Amherst}
  \state{MA}
  \country{USA}
}

\author{Jeremy Gummeson}
\email{jgummeso@umass.edu}
\orcid{0000-0002-7468-0569}
\affiliation{%
  \institution{University of Massachusetts Amherst}
  \city{Amherst}
  \state{MA}
  \country{USA}
}








\renewcommand{\shortauthors}{Abdullah Ahmed and Jeremy Gummeson}

\begin{abstract}
Latent spaces offer an efficient and effective means of summarizing data while implicitly preserving meta-information through relational encoding. We leverage these meta-embeddings to develop a modality-agnostic, unified encoder. Our method employs sensor-latent fusion to analyze and correlate multimodal physiological signals. Using a compressed sensing approach with autoencoder-based latent space fusion, we address the computational challenges of biosignal analysis on resource-constrained devices. Experimental results show that our unified encoder is significantly faster, lighter, and more scalable than modality-specific alternatives, without compromising representational accuracy.

\end{abstract}





\maketitle

\section{Introduction}

Modern machine learning systems depend on feature extraction to convert raw data into structured representations that capture semantically meaningful patterns \cite{Bengio2013}. By projecting high-dimensional inputs such as images or text into compact embedding spaces, these techniques enable efficient processing of salient features while filtering out noise \cite{LeCun2015}. For instance, convolutional neural networks (CNNs) learn hierarchical filters to isolate visual features like edges and textures \cite{He2016}, while word embeddings map linguistic tokens to vectors encoding semantic relationships \cite{Mikolov2013}. These representations make data machine-interpretable and support downstream tasks like similarity search and retrieval \cite{Pennington2014}. Recommendation systems use embeddings to model preferences \cite{Koren2009}, and retrieval-augmented language models employ dense vector indices for context-aware generation \cite{Lewis2020}. Such applications highlight how embedding spaces serve as a computational foundation for modern AI.

Deep learning has largely automated feature extraction, minimizing the need for manual engineering. CNNs, for example, are trained end-to-end to prioritize task-relevant features in domains ranging from medical imaging \cite{Litjens2017} to anomaly detection in sensor data \cite{Chandola2009}. A key development is the ability to unify embeddings across modalities: by encoding audio, video, and text into a shared space, cross-modal reasoning becomes possible \cite{Baltrusaitis2019}. CLIP \cite{Radford2021}, for example, aligns image-text pairs for zero-shot classification via contrastive learning. Similarly, multimodal transformers integrate diverse embeddings for tasks like video captioning \cite{Vaswani2017}. However, these systems typically rely on modality-specific encoders—such as ResNet-50 for images \cite{Russakovsky2015} or BERT for text \cite{Devlin2019}—which require considerable computational resources. For memory- and power-constrained devices like smartwatches or AR glasses, maintaining multiple specialized encoders for motion, voice, and biometric inputs is often infeasible.

\begin{figure}[htbp] \centering \includegraphics[width=0.8\linewidth]{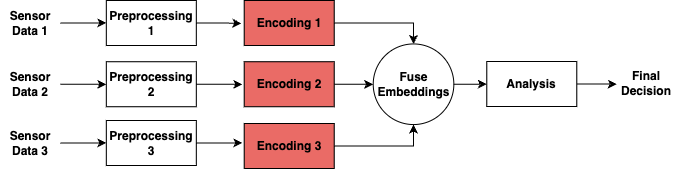} \caption{Traditional multimodal architectures rely on modality-specific encoders, increasing storage and compute demands for sensor data processing.} \label{fig:sysdesoverviewstandard} \end{figure}

To address this challenge, we explore shared latent spaces for efficient, modality-agnostic encoding. By leveraging the compression and information-preserving properties of latent representations \cite{DBLPconficlrHigginsMPBGBML17}, we aim to enable accurate and lightweight inference through a unified, information-rich encoding. Specifically, we evaluate vector-quantized variational autoencoders (VQ-VAEs) \cite{Oord2017} for multimodal information retrieval. Trained on generic image datasets, the encoder is designed to summarize visual patterns efficiently, regardless of the original signal modality. This unified feature space supports real-time applications on edge devices such as fitness trackers, where concurrent processing of PPG signals, accelerometer data, and voice input is essential for context-aware health monitoring \cite{10.48550/arxiv.2302.13155}.

Unlike conventional architectures (Figure \ref{fig:sysdesoverviewstandard}) that require dedicated encoders per modality, our system learns a unified latent space via a shared codebook. This approach improves scalability and reduces model complexity by 64\% relative to traditional sensor fusion frameworks \cite{10.48550/arxiv.2205.13542}. It delivers 1.4× faster inference while maintaining accuracy in cross-modal tasks. Our contribution demonstrates that lightweight, general-purpose embedding frameworks can meet performance demands, advancing the field of edge AI and enabling multimodal inference on constrained devices.

\begin{figure*}[htbp]
    \centering
    \includegraphics[width=0.85\linewidth]{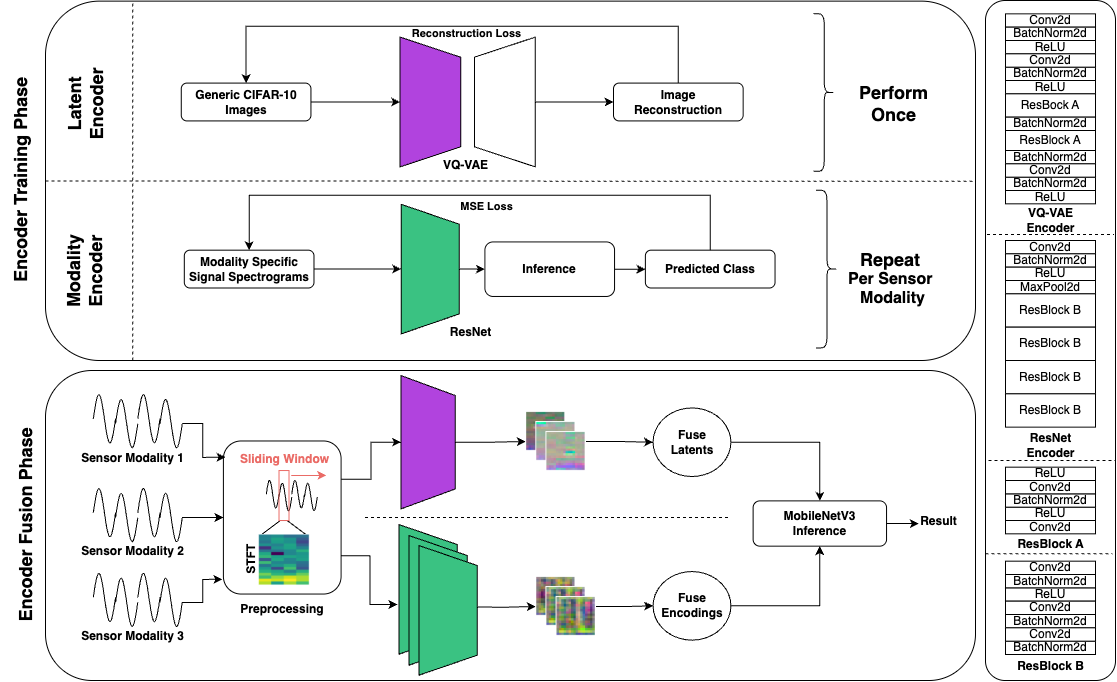}
    \caption{Complete sensor fusion pipeline, from encoder training to inference. Unlike traditional modality-specific encoders, our latent encoder is trained once on generic CIFAR-10 images, without exposure to modality-specific data. During inference, raw physiological signals undergo a preprocessing abstraction step, where they are transformed into spectral visualizations for encoding.}
    \label{fig:sysdesoverview}
\end{figure*}

\section{Related Work}\label{section:relatedwork}
BEVFusion \cite{10.48550/arxiv.2205.13542} integrates multi-LiDAR and multi-camera feeds through dual encoding stages within a shared Bird’s Eye View (BEV) space. Each sensor stream is first processed by a dedicated encoder to extract modality-specific features, which are then fused via a BEV encoder. While the system achieves comparable accuracy and reduces computation by up to 1.9×, its architecture remains constrained by its reliance on multiple specialized encoders.

SparseFusion \cite{10.1109/iccv51070.2023.01613} addresses noise in BEV-based representations by applying sparse sampling across modalities. This technique focuses on partial signal segments, enabling efficient generation of less dense 3D object representations from high-resolution LiDAR and camera data. Similarly, FreqMAE \cite{10.1145/3589334.3645346} employs masked autoencoders to learn semantic features from IoT sensor signals via partial reconstruction. This enables self-supervised learning without extensive labeling or data augmentation. However, these systems still depend on modality-specific encoders, which limit scalability and generalization. While FreqMAE adopts autoencoders for unsupervised feature extraction, its encoders are trained to capture modality-specific structure and trained on using modality-specific datasets, rather than generic visual patterns as proposed in our unified encoding approach (Section \ref{section:unifieddataembedding}).

Most similar to our method are Multimodal Variational Autoencoders (MVAEs), which leverage generative VAEs \cite{10.48550/arxiv.1906.02691} to embed multimodal data into a shared latent space \cite{10.48550/arxiv.1802.05335}. These models employ a two-stage architecture: modality-specific encoders extract features before projecting them into a common, semantically rich latent representation. Despite their success in aligning heterogeneous data, MVAEs still require a distinct encoder per modality to accommodate differing input structures.

In contrast, we propose a modality-agnostic feature extractor based on a VQ-VAE combined with a lightweight preprocessing step that spectrally visualizes all input signals. This enables training a single encoder on generic image datasets (e.g., CIFAR-10) while retaining the ability to process diverse physiological modalities. Unlike MVAE, our system substitutes generative modeling with a simpler inference task, significantly reducing training data requirements and computational overhead. The full architecture and design rationale are detailed in Section \ref{chapter:systemdesign}.

\section{Methods}\label{chapter:systemdesign}
To enable computationally efficient sensor fusion on mobile devices, we propose a unified encoder-based pipeline that minimizes processing overhead. The pipeline begins with a preprocessing stage that abstracts raw data streams into a common visual format compatible with a generic image-based semantic encoder. This encoder transforms inputs into a unified, semantically rich, and memory-efficient embedding space. These embeddings are then directly fused and analyzed by task-specific machine learning (ML) models. Figure \ref{fig:sysdesoverview} illustrates the full system pipeline from data ingestion to fusion.

\subsection{Resampling and Feature Extraction}\label{subsection:dataresampling}
As discussed, visualizing time-series signals as images allows us to use a single encoder across modalities. This preprocessing step abstracts raw signals into spectral images of uniform dimensions and sampling rates. We apply forward-filling to maintain consistency during deployment and use sliding window resampling (128-record windows with a 32-step overlap) to account for temporal variability. Segmented time-series data are converted into spectral images using the Short-Time Fourier Transform (STFT):

\begin{equation} X(\tau, \omega)=\int_{-\infty}^{\infty} x(t) w(t-\tau) e^{-i \omega t} d t \label{eq:stfteq} \end{equation}

This transformation produces standardized 128×128 RGB images \cite{10.1109/jproc.2018.2820126}, enabling compatibility with image-based encoders while preserving temporal and frequency content. Residual gaps are handled through zero-filling to ensure continuity \cite{10.1109/23.589532}. The resulting representations retain physiologically relevant features while supporting real-time processing.

\subsection{Unified Data Embedding}\label{section:unifieddataembedding}
We adopt a Vector Quantized Variational Autoencoder (VQ-VAE) \cite{10.48550/arxiv.1711.00937} to encode spectral images of multimodal sensor data into a unified latent space. The encoder-decoder pair is trained using the Evidence Lower Bound (ELBO) objective:

\begin{equation} \mathcal{L}{\text{ELBO}} = \mathbb{E}{q(z|x)}[\log p(x|z)] - \beta D_{\text{KL}}(q(z|x) \parallel p(z)) \end{equation}

This formulation mitigates common VAE issues such as posterior collapse and latent entanglement \cite{10.48550/arxiv.2012.09841}, while enabling the following capabilities:

\begin{itemize} \item \textbf{Modality-agnostic processing} via spectral abstraction into a common image space. \item \textbf{Structured latent representations} that promote interpretability and downstream compatibility. \item \textbf{Cross-modal feature alignment} through discrete codebook-based encoding. \end{itemize}

The trained encoder is then frozen to produce compact representations, mapping 128×128 RGB spectral images into 16×16 latent codes. These embeddings capture modality-invariant physiological patterns while supporting efficient, real-time inference.

\section{Experiments}
We evaluate our unified latent encoder using a VQ-VAE model originally pretrained to reconstruct generic images. During inference, only the encoder is used as part of our embedding-based stress classification pipeline. The VQ-VAE architecture comprises convolutional and residual blocks. To assess the effectiveness of the learned latent space, we fine-tune a Conv-LSTM network for binary classification of psychological stress levels.

As a baseline, we compare our method to a BEV-fusion-inspired approach \cite{10.48550/arxiv.2205.13542}, using modality-specific encoders constructed from fine-tuned Residual Networks (ResNet) \cite{10.48550/arxiv.1512.03385}. For each modality, we train a ResNet encoder on 1,000 labeled samples, then splice the ingestion layers to serve as encoders for the fusion architecture.

\subsection{Implementation}\label{section:encoderimplementation}
We implement and test our latent encoder using data from the WESAD dataset \cite{10.1145/3242969.3242985}, which includes high-frequency chest-mounted RespiBAN recordings and psychometric surveys (PANAS, STAI) from 15 participants over two hours. We fuse six physiological signals: electrocardiogram (ECG), electromyography (EMG), electrodermal activity (EDA), skin temperature, respiration rate, and 3-axis accelerometer data (Acc), according to the permutations in Table \ref{tab:modality}.
All experiments are conducted using PyTorch 2.3 on an M1 Pro CPU. Standard preprocessing steps include timestamp conversion, binary gender encoding, and numerical BMI/age transformation. Window-sampled signals (3×128×128 RGB) are encoded into compact 3×16×16 latent representations using a VQ-VAE pretrained on CIFAR-10 \cite{10.48550/arxiv.1711.00937}. 
For downstream inference, we fine-tune a MobileNetV3 model \cite{10.48550/arxiv.1905.02244} with Conv-LSTM layers to capture temporal dynamics. Final layers classify each sequence as high or low stress based on reported STAI scores, demonstrating the feasibility of efficient physiological computing in mobile environments.

\subsection{Unified Compression Runtime Performance}

\begin{figure}[htbp] 
\centering 
\includegraphics[width=1\linewidth]{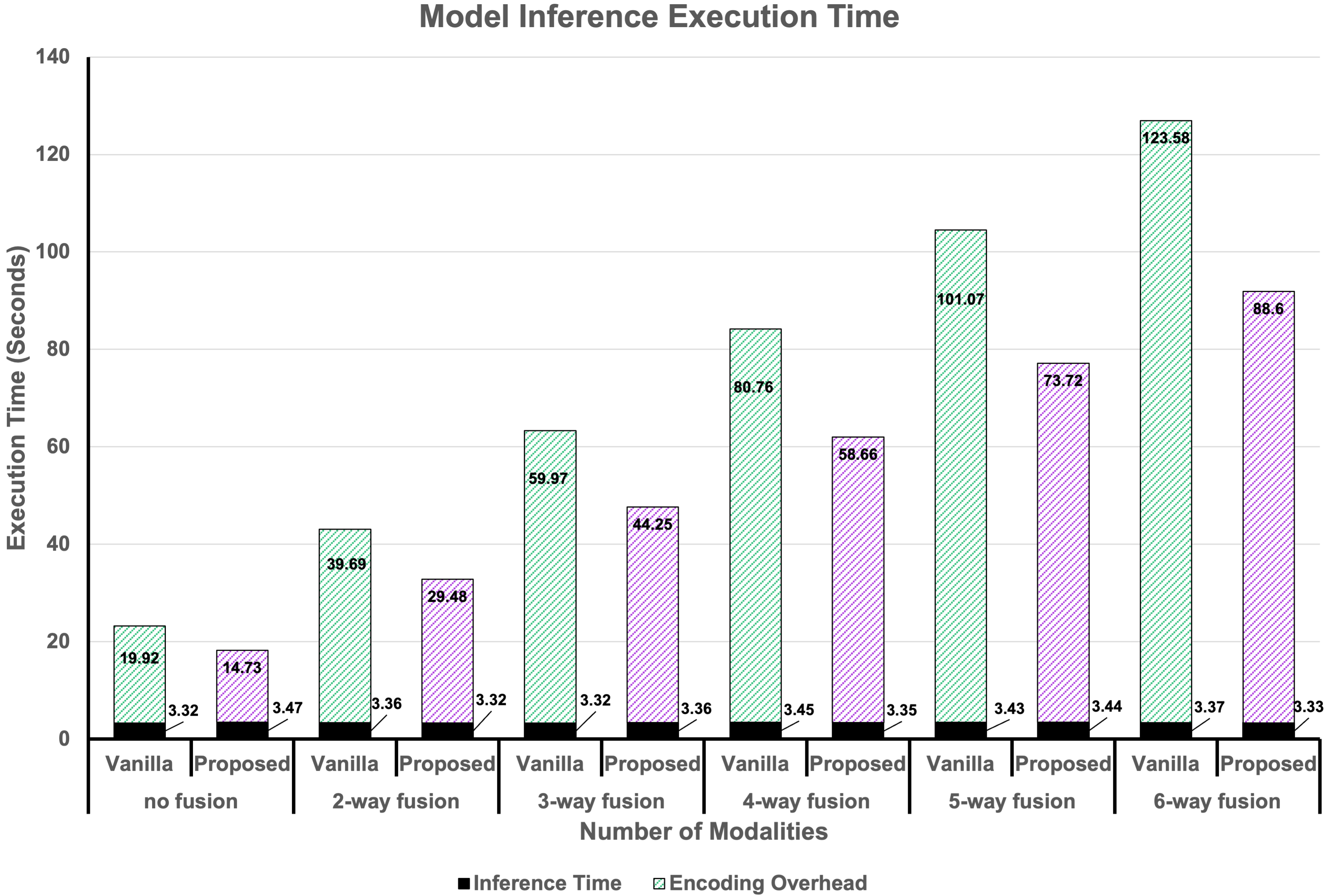} 
\caption{Inference runtime, highlighting encoding overhead. As the number of modalities increases, the performance gap widens, demonstrating the runtime benefits of a generic latent encoder.} 
\label{fig:inferencecomputetime} 
\end{figure}

To evaluate encoding overhead, we conducted controlled runtime experiments using pretrained inference models (Figures \ref{fig:accuracy}–\ref{fig:runtimecomplexity}). Latency measurements isolate the encoder’s contribution during inference. In simplified tasks, this overhead is more pronounced, partly offsetting the benefits of efficient decoding. However, in more complex tasks where inference dominates total runtime, our encoder’s efficiency becomes significantly advantageous. Further optimizations to the VQ-VAE architecture offer potential for even greater runtime reductions.

\begin{table}[htbp]
  \caption{Legend of Modality Fusion Permutations}
  \label{tab:modality}
  \begin{tabular}{cl}
    \toprule
   Number of Modalities & List of Modalities Fused\\
    \midrule
    1& ECG\\
    2& ECG, EMG\\
    3& ECG, EMG, EDA\\
    4& ECG, EMG, EDA, Temp\\
    5& ECG, EMG, EDA, Temp, Resp\\
    6& ECG, EMG, EDA, Temp, Resp, Acc\\
    \bottomrule
  \end{tabular}
\end{table}

\subsection{Unity of Representation}
\label{section:unityofrepresentation}
\begin{figure}[htbp] 
\centering 
\includegraphics[width=1\linewidth]{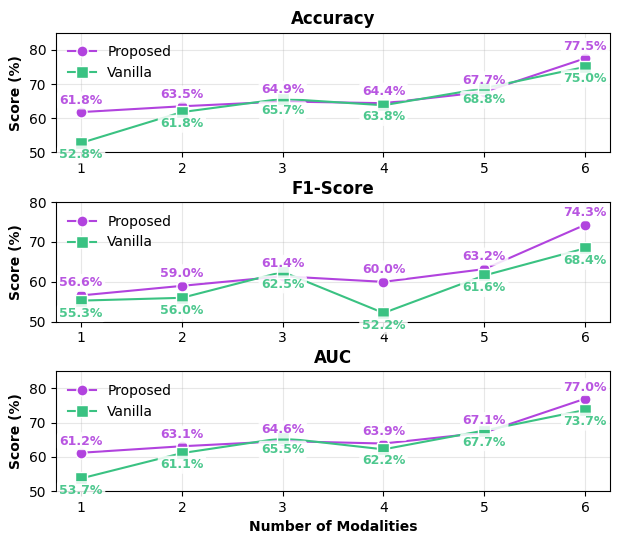} \caption{Performance across increasing modality fusion count. Latent encoding yields stable and accurate classification, even as signal diversity increases.} 
\label{fig:accuracy} 
\end{figure}

Our evaluation demonstrates that a unified image encoder can support effective cross-modal integration through shared latent representations. Using the modality permutations in Table \ref{tab:modality}, we fuse up to six physiological signals. As shown in Figure \ref{fig:accuracy}, our method consistently matches or exceeds BEV-fusion in accuracy, F1-score, and AUC metrics, validating the encoder’s modality-agnostic capability. Notably, our model exhibits greater stability as the number of modalities increases. While all metrics dip slightly upon the inclusion of the fourth modality (skin temperature)—likely due to its limited relevance to stress prediction—our method recovers more quickly, suggesting improved robustness and potential noise tolerance in the latent space.

\subsection{Memory and Power Analysis}

\begin{figure}[htbp]
    \centering
    \includegraphics[width=0.8\linewidth]{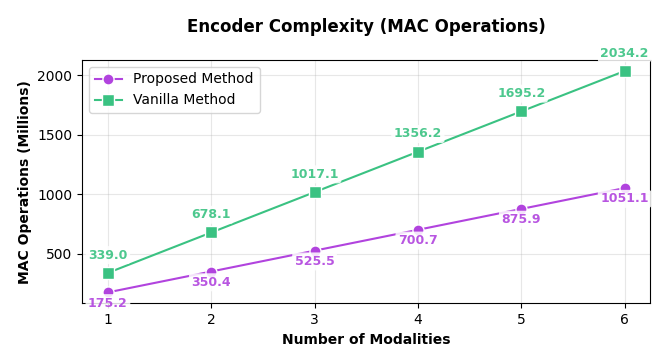}
    \caption{Encoder complexity measured in Multiply-Accumulate (MAC) operations. Although complexity scales linearly for both methods, the vanilla approach exhibits a steeper growth rate, resulting in significantly higher computational demands as sensor count increases.}
    \label{fig:runtimecomplexity}
\end{figure}

We assess resource efficiency by estimating model complexity (in MAC operations) and memory requirements via fetch-write activity. As shown in Figure \ref{fig:runtimecomplexity}, our latent encoder is 1.9× less computationally complex than BEV-fusion and requires less memory even when compared to a single BEV-fusion encoder. Unlike modality-specific approaches that scale linearly with the number of sensors, our method requires loading only one encoder regardless of modality count. This yields consistent memory usage and reduces approximate energy consumption, underscoring its suitability for deployment on edge devices.

\section{Limitations and Future Work}
While this work highlights the latent space's potential for modality-agnostic fusion and noise tolerance, further experimentation is necessary to fully assess its generalizability. A primary limitation is the scope of evaluation—our unified encoding scheme was tested on a limited number of modalities, datasets, and tasks. Future work will expand testing to datasets with more diverse sensor types to evaluate performance under more demanding fusion scenarios.
We also plan to investigate the role of peak signal-to-noise ratio (PSNR) in influencing information retention and inference quality, particularly in low-quality or sparse input settings. This analysis will help determine the extent to which latent-space fusion can mitigate sparsity. Lastly, we aim to deploy and profile our model on actual edge devices to empirically assess energy consumption and runtime efficiency in real-world conditions.

\section{Conclusion}
This work introduces a unified encoder design that reduces computational, memory, and energy requirements while enabling efficient multimodal data fusion. We demonstrate that modality-agnostic encoding can preserve sufficient semantic information for direct inference, eliminating the need to reconstruct input signals in their original format. 
Future work will explore feature selection and ranking methods to better understand the versatility and limits of latent fusion. We also aim to investigate the robustness of latent codes to real-world noise and sparsity, as well as explore alternative fusion strategies to further optimize performance for edge-based multimodal systems.
We believe continued research into encoder efficiency, latent-space inference, and compact representation learning can help advance the development of resilient, scalable, and real-time systems for on-device informatics.

\bibliographystyle{ACM-Reference-Format}
\balance
\bibliography{AcceptedEditedV2}

\end{document}